\documentstyle[preprint,aps]{revtex}
\begin{document}

\tightenlines

\draft


\title{ Quantization of the Nonlinear Schr\"odinger Equation \break
        on the Half Line }
\author{ Mario Gattobigio}
\address{  Istituto Nazionale di Fisica Nucleare, Sezione di Pisa,
Dipartimento di Fisica dell'Universit\`a di Pisa
Piazza Torricelli 2, 56100 Pisa, Italy}
\author{  Antonio Liguori}
\address{ International School for Advanced Studies, 34014 Trieste, Italy}
\author{ Mihail Mintchev}
\address{  Istituto Nazionale di Fisica Nucleare, Sezione di Pisa,
Dipartimento di Fisica dell'Universit\`a di Pisa
Piazza Torricelli 2, 56100 Pisa, Italy}

\maketitle
\begin{abstract}
We establish the second quantized solution of
the nonlinear Schr\"odinger equation on the half line
with a mixed boundary condition. The solution is based on
a new algebraic structure, which we call boundary exchange
algebra and which substitutes, in the presence of boundaries,
the familiar Zamolodchikov-Faddeev algebra.
\end{abstract}




The recent interest in quantum field theory on
the half line ${\bf R}_+ = \{x\in{\bf R} : x>0\}$
is related to some successful applications
in open string theory, dissipative quantum mechanics and
impurity problems in condensed matter physics.
This paper concerns the quantization of the
nonlinear Schr\"odinger equation (NLS)
\begin{equation}
(i\partial_t + \partial_x^2 )\Phi (t,x) = 2g\, |\Phi (t,x)|^2 \Phi (t,x)
\; , \; \; \; g > 0 \; ,
\end{equation}
on ${\bf R}_+$ with the boundary condition
\begin{equation}
\lim_{x \downarrow 0} ( \partial_x - \eta )
\Phi (t,x) = 0 \; , \; \; \;  \eta \geq 0\; .
\end{equation}
Our main result is the exact second quantized solution of
the boundary valued problem (1,2).

When considered on the whole line ($x\in \bf R$), Eq.(1) gives
rise to one of the most popular and extensively studied integrable
system, which has been solved [1-4] by the method of
inverse scattering transform. Let us summarize briefly those
main results in $\bf R$, which turn out to be relevant for our
investigation in ${\bf R}_+$.
A convenient starting point is a classical
solution of Eq.(1), discovered time ago by Rosales [5] and
reading
\begin{equation}
   \Phi(t,x) = \sum_{n=0}^{\infty} (-g)^n  \Phi^{(n)}(t,x)
\; ,
\end{equation}
\begin{eqnarray}
 \Phi^{(n)} && (t,x) =
\int \prod_{i=1\atop j=0}^n {dp_i \over 2\pi}{dq_j \over 2\pi}
{\overline \lambda }(p_1)\cdots {\overline \lambda }(p_n)
\, \lambda (q_n)\cdots \lambda (q_0) \nonumber \\
&& \qquad
\times { {\rm e}^{ i\sum_{j=0}^n (xq_j-tq^2_j) -i\sum_{i=1}^n (xp_i-tp^2_i) }
\over \prod_{i=1}^n \left[ (p_i - q_{i-1})\, (p_i - q_{i}) \right]}
\; .
\end{eqnarray}
The bar indicates complex conjugation and the integration is
defined by the principal value prescription. It is assumed that
the function $\lambda $ is such that
the integrals in (4) exist and the series (3)
converges for sufficiently small $g$. Any Schwartz test function
meets for instance these requirements. A remarkable property of
the solution (3,4) is that its general structure is preserved
by the quantization. Indeed, following [2-4], the quantum
solution on $\bf R$ admits the series expansion (3) and
the $n$-th order contribution has a similar form, namely
\begin{eqnarray}
 &&  \Phi^{(n)}(t,x) =
\int \prod_{i=1\atop j=0}^n {dp_i \over 2\pi}{dq_j \over 2\pi}
a^*(p_1)\cdots a^*(p_n) \, a(q_n)\cdots a(q_0) \nonumber \\
&& \qquad
\times { {\rm e}^{ i\sum_{j=0}^n (xq_j-tq^2_j) -i\sum_{i=1}^n (xp_i-tp^2_i) }
\over \prod_{i=1}^n \left[ (p_i - q_{i-1} - i\epsilon )\,
 (p_i - q_{i} - i\epsilon ) \right]}
\; .
\end{eqnarray}
Besides the $i\epsilon $ prescription to contour poles, the new
and fundamental feature is the substitution of the the functions
$\{\lambda (k) , {\overline \lambda }(k)\}$ with
the generators $\{a(k) , a^*(k)\}$ of the
Zamolodchikov-Faddeev (ZF) [6] algebra ${\cal A}_R$ defined by
\begin{eqnarray}
&& a(k_1) \, a(k_2)  \, -  R(k_2,k_1) a(k_2) \, a(k_1)    = 0 \;  ,
\nonumber \\
&& a^*(k_1) a^*(k_2)    -  R(k_2,k_1) a^*(k_2) a^*(k_1)   = 0 \;  , \\
&& a(k_1) \, a^*(k_2)   -  R(k_1,k_2) a^*(k_2) \, a(k_1)  =
2\pi \delta (k_1-k_2) \; , \nonumber
\end{eqnarray}
where
\begin{equation}
R(p,k)={p-k-ig \over p-k+ig}
\; .
\end{equation}
In Eq.(5) $\{a(k) , a^*(k)\}$ are taken in the Fock representation
(see e.g. [7,8]) ${\cal F}_R$ of ${\cal A}_R$, which provides therefore
the state space of the quantum NLS. As usual, the $*$-operation is
realized as Hermitian conjugation with respect to the scalar
product in ${\cal F}_R$. In order to give a precise meaning of
the cubic term in (1), one introduces a normal ordering $: ... :$
between $a,a^*$ by putting all annihilation operators $a$
to the right and all creation operators $a^*$ to the left, conserving
separately the original order among creators and annihilators themselves.
The quantum version of Eq.(1) is then obtained by the substitution
\begin{equation}
|\Phi (t,x)|^2 \Phi (t,x)\, \mapsto \; : \Phi \Phi^* \Phi : (t,x)
\; ,
\end{equation}
and can be verified explicitly [4]. An essential point is also that
the equal-time canonical commutation relations
\begin{eqnarray}
&& [\Phi(t,x),\Phi(t,y)]= [\Phi^*(t,x),\Phi^*(t,y)]= 0 \; , \\
&&[\Phi(t,x),\Phi^*(t,y)]= \delta(x-y) \; ,
\end{eqnarray}
hold [4] on a dense domain in ${\cal F}_R$.

Eq.(5) is the quantum inverse scattering transform for the
NLS on $\bf R$. It allows to reconstruct the off-shell
quantum field $\Phi (t,x)$ from the scattering data encoded in the Fock
representation ${\cal F}_R$ of ${\cal A}_R$. One can show [1-4] in fact,
that the relative asymptotic states can be represented by
\begin{eqnarray}
|k_1,...,k_n\rangle^{\rm in} &=& a^*(k_1)...a^*(k_n)\Omega
\qquad k_1>...>k_n \; ,  \\
|p_1,....,p_n\rangle^{\rm out} &=& a^*(p_1)...a^*(p_n)\Omega
\qquad  p_1<...<p_n \; ,
\end{eqnarray}
where $\Omega \in {\cal F}_R$ is the vacuum state and
$\{ k_1,...,k_n\}$ and $\{ p_1,...,p_n\}$ denote the
momenta of the incoming and outgoing particles respectively.
Varying $n \geq 0$, the linear subspaces generated by the vectors (11) and
(12) give rise to the corresponding asymptotic spaces ${\cal F}_R^{\rm in}$
and ${\cal F}_R^{\rm out}$, which turn out [8] to be separately
dense in ${\cal F}_R$. By means of Eqs.(6) one easily derives
the scattering amplitudes, which have the factorized form
\begin{eqnarray}
&& {}^{\rm out} \langle p_1,...,p_m|k_1,...,k_n \rangle ^{\rm in}=
\nonumber \\
\delta_{mn}\prod_{i=1}^n && 2\pi \delta(p_i-k_{n+1-i})
\prod_{i,j=1 \atop i<j}^n R(p_i,p_j)
\, .
\end{eqnarray}
As it can be expected from integrability, these amplitudes
vanish unless $n=m$. Moreover, the particle momenta are
separately conserved, $p_i = k_{n+1-i},\; i =1,...,n$. From
Eq.(13) we deduce also that the exchange factor $R$ in the ZF
algebra ${\cal A}_R$ is actually the two-body scattering matrix.

At this stage we have enough background for facing the
boundary valued problem (1,2) on the half line ${\bf R}_+$.
Its classical integrability has been investigated in [9].
After some algebra one can verify [9,10] that besides Eq.(1),
the field $\Phi (t,x)$ defined by Eqs.(3,4) solves also
the boundary condition (2), provided that
the function $\lambda (k)$ satisfies the reflection condition
\begin{equation}
         \lambda (k) = B(k) \lambda (-k) \; ,
\end{equation}
with reflection coefficient given by
\begin{equation}
B(k)={k-i\eta \over k +i \eta}
\; .
\end{equation}
So, as a first tentative to quantize Eqs.(1,2), one may try to
keep Eqs.(3,5) and to implement in ${\cal A}_R$ the
analog of condition (14), namely $a(k) = B(k)a (-k)$.
It is easily seen however that such a constraint
is not compatible with the exchange relations (6)
and the two-body bulk scattering matrix (7).

The next natural conjecture at this point is that the quantum
solution of Eqs.(1,2) is still of the form (3,5), but for
the replacement of ${\cal A}_R$ with another appropriate
algebraic structure ${\cal B}_R$, called in what follows boundary
algebra. It will be shown below that this conjecture is right
and the main problem is to determine ${\cal B}_R$, finding
in particular the quantum counterpart of condition (14).
Our strategy in analyzing this problem will be as follows.
We will first recall the scattering
amplitudes corresponding to the two-body bulk scattering matrix
(7) and the reflection coefficient (15). From these amplitudes
one can recover the underlying
boundary algebra ${\cal B}_R$. The final step will be
to check that Eq.(5) in the Fock representation ${\cal F}_{R,B}$ of
${\cal B}_R$ provides the solution of (1,2,9,10).

The scattering theory of one-dimensional integrable
systems in the presence of a reflecting boundary has
been developed by Cherednik [11] and successfully applied
more recently by Ghoshal and Zamolodchikov [12]. The following
picture emerges from these investigations.
Let $|k_1,...,k_n \rangle^{\rm in}$ be an in-state,
representing $n$ particles coming from $x=+\infty$ and
thus having negative momenta $k_1<k_2<...<k_n<0$.
These particles interact among themselves before and after being
reflected by the wall at $x=0$, giving rise to
an out-state $|p_1,...,p_m \rangle^{\rm out}$
composed of particles traveling toward $x=+\infty$ and thus having
positive momenta \break $p_1>p_2>...>p_m>0$.
The transition amplitude between these states vanishes unless
$n=m$ and $p_i = -k_i$, $i=1,...,n$. Therefore, not only
the total momentum, but each momentum is separately reflected.
According to [11], the scattering amplitude is
\begin{eqnarray}
&&  {}^{\rm out} \langle p_1,...,p_m |k_1,...,k_n \rangle ^{\rm in} =
\nonumber \\
\delta_{mn} \prod_{i=1}^n 2\pi \delta(p_i && +k_{i})B(p_i)
\prod_{i,j=1 \atop i<j}^n R(p_i,p_j)R(p_i,-p_j) \; .
\end{eqnarray}
The $R$-factors describe the interactions among the
particles in the bulk, while the $B$-factors take into account the
reflection from the wall. The crucial point now is that
there exists an algebra ${\cal B}_R$ encoding the
scattering amplitudes (16), in perfect analogy to
the case without boundary, where (13) are related to
${\cal A}_R$. ${\cal B}_R$ is the boundary algebra we
are looking for. It plays the main role in the quantum
inverse scattering approach to the boundary valued
problem (1,2) and has been introduced and
investigated in a more general context in [13]. In the specific
case under consideration, ${\cal B}_R$ is generated by $\{a(k),a^*(k),b(k)\}$.
These generators satisfy quadratic exchange relations,
which can be conveniently grouped in two sets.
\vfill
\eject
The first one is
\begin{eqnarray}
a(k_1) \, a(k_2)  \, &-&  R(k_2,k_1) a(k_2) \, a(k_1)    = 0 \;  ,\nonumber \\
a^*(k_1) a^*(k_2)    &-&  R(k_2,k_1) a^*(k_2) a^*(k_1)   = 0 \;  , \\
a(k_1) \, a^*(k_2)   &-&  R(k_1,k_2) a^*(k_2) \, a(k_1)  =
2\pi \delta (k_1-k_2) \nonumber \\
  &+& b(k_1)  2\pi
\delta (k_1+k_2) \; , \nonumber
\end{eqnarray}
and strongly resembles (6), except for the presence of the so called
boundary generator $b(k)$ in the right hand side of the last equation.
The second set of constraints describes
the exchange relations of $b(k)$ and reads
\begin{eqnarray}
a(k_1) b(k_2)   & =& R(k_2,k_1) R(k_1,-k_2) \,  b(k_2) a(k_1) \; ,
\nonumber \\
b(k_2) a^*(k_1) & =& R(k_2,k_1) R(k_1,-k_2) \,  a^*(k_1)b(k_2) \; , \\
b(k_1)b(k_2)    &= & b(k_2)b(k_1) \; . \nonumber
\end{eqnarray}
We observe that the coupling constant $g$ enters the algebra
directly through the exchange factor $R$, while there is still no
reference to the boundary parameter $\eta $: it determines which of
the inequivalent Fock representations of ${\cal B}_R$
must be chosen. Indeed, as explained in details in [13], a Fock
representation ${\cal F}_{R,B}$ of ${\cal B}_R$ is characterized
by a vacuum state $\Omega $, such that:
\begin{enumerate}
\item {$\Omega $ is annihilated by $a(k)$;}
\item {$\Omega $ is cyclic with respect $a^*(k)$;}
\item {$\Omega $ is an eigenvector of $b(k)$ with eigenvalue $B(k)$.}
\end{enumerate}
One must distinguish the c-number reflection coefficient $B(k)$ from
the boundary generator $b(k)$, which according to
Eqs.(18) does not even commute with $\{a(k), a^*(k)\}$.

Recapitulating, the mere fact that our system has a
boundary shows up at the algebraic level,
turning the ZF algebra ${\cal A}_R$ into the boundary algebra
${\cal B}_R$. According to point 3 above, the details of the
boundary condition (2) (namely, the value of the parameter $\eta $)
enter at the representation level through the reflection coefficient
$B(k)$. In the Fock space ${\cal F}_{R,B}$ one has [13]
\begin{equation}
         a(k) = b(k) a(-k) \; ,
\end{equation}
which is the correct quantum analogue of Eq.(14) and descends from a
peculiar automorphism of ${\cal B}_R$.

The connection with Cherednik's scattering theory is obtained
through the identification
\begin{eqnarray}
|k_1,...,k_n\rangle^{\rm in} &= a^*(k_1)...a^*(k_n)\Omega \; ,
\quad  k_1<...<k_n<0 \; , \nonumber \\
|p_1,...,p_n\rangle^{\rm out} &= a^*(p_1)...a^*(p_n)\Omega \; ,
\quad  p_1>...>p_n>0 \; \nonumber .
\end{eqnarray}
One finds in fact, that the amplitudes
\[
(a^*(p_1)...a^*(p_m)\Omega \, ,\, a^*(k_1)...a^*(k_n)\Omega ) \; ,
\nonumber
\]
where $(\cdot , \cdot )$ is the scalar product in ${\cal F}_{R,B}$,
precisely reproduce the right hand side of Eq.(16).

The final step of our consideration is to show that
inserting in Eq.(5) the generators $\{a(k), a^*(k)\}$
in the Fock representation ${\cal F}_{R,B}$ of ${\cal B}_R$
and restricting to $x>0$, one gets a field $\Phi (t,x)$
which solves Eqs.(1,2) and satisfies the commutation relations
(9,10). The details in verifying the validity of this statement
are given in [10]. Here we shall focus on the essential points
only. First of all, we would like to fix
an appropriate domain ${\cal D} \subset {\cal F}_{R,B}$ for
the quantum fields. For this purpose we denote by ${\cal D}^n$
with $n\geq 1$ the $n$-particle subspace
\[
\left \{ \int dp_1...dp_n f(p_1,...,p_n)a^*(p_1)...a^*(p_n)\Omega
\, :\, f\in {\cal S}({\bf R}^n) \right \}.
\]
Setting ${\cal D}^0 = \{v\Omega \, :\, v\in \bf C\}$ we then define
$\cal D$ as the linear space of sequences \break
$\varphi = (\varphi^{(0)}, \varphi^{(1)},...,\varphi^{(n)},...)$
with $\varphi^{(n)} \in {\cal D}^{(n)}$ and $\varphi^{(n)}=0$ for
$n$ large enough. The last condition and the normal ordered structure
of $\Phi^{(n)}(t,x)$ directly imply that the series (3) converges
in mean value on $\cal D$ for any $g>0$.
By cyclicity of the vacuum, $\cal D$ is dense
in ${\cal F}_{R,B}$ and one can show [10] that also Eqs.(1,2) are
satisfied in mean value on $\cal D$. In fact,
\begin{equation}
(i\partial_t + \partial_x^2 )(\varphi_1, \Phi(t,x) \varphi_2 )
= 2g (\varphi_1, :\Phi \Phi^* \Phi :(t,x) \varphi_2 ) \; ,
\end{equation}
\[
\lim_{x \downarrow 0} ( \partial_x - \eta )
(\varphi_1, \Phi(t,x) \varphi_2 ) = 0  \: , \nonumber
\]
hold for any $\varphi_1 , \varphi_2 \in {\cal D}$ and $x>0$.
The proof of Eq.(20) is similar in spirit to that given
by Davies [4] for the NLS on $\bf R$. The novelty consists in
evaluating the contributions of the boundary generator $b$
stemming from the exchange of $a$ with $a^*$. In this respect
the conditions $x>0$ and $\eta \geq 0$ show to be essential.

As it should be expected, the fields $\Phi (t,x)$ and
$\Phi^*(t,x)$ have to be regarded as sesquilinear forms on
$\cal D$. In order to deal with operators, one has to take
the averages
\[
\Phi(t,h) = \int dx \Phi (t,x)h(x) ,  \; \;
\Phi^*(t,h) = \int dx \Phi^* (t,x)h(x) ,
\]
with $h\in {\cal S}({\bf R})$ and such that
${\rm supp}h \subset {\overline {\bf R}}_+$.
One has by construction
\[
\Phi (t,h){\cal D}^{(0)} = 0 \; , \qquad
\Phi (t,h){\cal D}^{(n)} \rightarrow {\cal D}^{(n-1)} \; ,
\quad n \geq 1 \; , \nonumber
\]
\[
\Phi^* (t,h){\cal D}^{(n)} \rightarrow {\cal D}^{(n+1)} \; ,
\quad n \geq 0 \; . \nonumber
\]
$\cal D$ is a common invariant domain for the operators $\Phi(t,h)$
and $\Phi^*(t,h)$, where the equal-time commutation relations
\begin{eqnarray}
&& [\Phi(t,h_1),\Phi(t,h_2)]= [\Phi^*(t,h_1),\Phi^*(t,h_2)]= 0 \; , \\
&& [\Phi(t,h_1),\Phi^*(t,h_2)]= \int dx h_1(x) h_2(x)  \; ,
\end{eqnarray}
hold [10]. Moreover, the vacuum $\Omega $ is cyclic also with
respect to $\Phi^* (t,h)$ for fixed $t$.

Concerning the time-evolution, one can define on $\cal D$
the essentially self adjoint operator
\begin{equation}
H = \int {dp \over 2\pi}{p^2\over 2}  a^*(p) a(p)  \; ,
\end{equation}
which generates
\[
{\rm e}^{i H t} a(k) {\rm e}^{-iHt} = {\rm e}^{-i k^2 t} a(k) \; ,
\quad
{\rm e}^{i H t} b(k) {\rm e}^{-iHt} = b(k) \; .
\]
Now, it is easily seen that
\[
\Phi(t,h) =  {\rm e}^{i H t} \Phi(0,h) {\rm e}^{-iHt} \; , \nonumber
\]
which shows that $H$ is actually the Hamiltonian of the NLS model on
the half line. Notice the simple form (23) of $H$ in terms of $\{a(k), a^*(k)\}$
and the fact that $b(k)$ does not evolve in time.

The nontrivial correlation functions of our system involve
equal number of $\Phi $ and $\Phi^*$. From the structure of
Eq.(5) it follows that for computing the exact $2n$-point
function one does not need all terms in the expansion (3), but at
most the $(n-1)$-th order contribution. For instance, the
two-point function is given by
\begin{eqnarray}
&&(\Omega \, ,\, \Phi(t_1,x_1)\Phi^*(t_2,x_2)\Omega ) = \nonumber \\
=\int {dp \over 2\pi} &&{\rm e}^{-ip^2 (t_1-t_2)}
\left [ {\rm e}^{ip(x_1-x_2)}+B(p) {\rm e}^{ip(x_1+x_2)} \right ] \; ,
\end{eqnarray}
and coincides with that of the non-relativistic free field on
half line. In that context $\eta $ parametrizes
a family of self-adjoint extensions
of the Laplacian on ${\bf R}_+$. The nontrivial scattering is consistently
described by the $2n$-point correlation functions for $n\geq 2$.
These functions differ from the free ones and their on-shell limit
leads [10] to the transition amplitudes (16), which completes the picture
and concludes our quantum field theory description of the NLS model on
${\bf R}_+$.

Summarizing, we have shown above that the quantum inverse scattering
transform (5) works for solving the boundary valued problem (1,2),
provided that the ZF algebra ${\cal A}_R$ is replaced by the
boundary algebra ${\cal B}_R$. It will be interesting to extend
our result to the case $\eta < 0$, where the presence of boundary bound
states must be taken into account.

\end{document}